\documentclass[final,authoryear,5p,times,twocolumn]{elsarticle}

\usepackage{graphicx}
\usepackage{epsfig}
\usepackage{amssymb}
\usepackage{amsthm}
\usepackage{graphicx}
\usepackage[fleqn]{amsmath}
\usepackage{txfonts}
\usepackage{algorithm,algorithmic}
\usepackage{subfigure}
\usepackage{color}

\def\INTEGRALSPI{\textit{INTEGRAL/SPI}}
\def\INTEGRAL{\textit{INTEGRAL}}

\def\SPI{SPI}

\def\timebins{\textquotedblleft{}time-bins\textquotedblright{}}

\def\apj{ApJ} 
\def\mnras{MNRAS} 
\def\aap{A\&A} 

\journal{Astronomy and Computing}

\begin{document}

\begin{frontmatter}

\title{Simultaneous analysis of large \INTEGRALSPI\tnoteref{labela} datasets:
optimizing the computation of the solution and its variance using
sparse matrix algorithms}

\tnotetext[labela]{Based on observations with INTEGRAL, an ESA project
with instruments and science data center funded by ESA member states
(especially the PI countries: Denmark, France, Germany, Italy, Spain,
and Switzerland), Czech Republic and Poland with participation of
Russia and the USA.}

\author[inst1,inst2]{L.~Bouchet\corref{cor1}}
\author[inst3]{P.~Amestoy}
\author[inst4]{A.~Buttari}
\author[inst3,inst5]{F.-H.~Rouet}
\author[inst1,inst2]{M.~Chauvin}

\address[inst1]{Universit\'e de Toulouse, UPS-OMP, IRAP, Toulouse,
France} 
\address[inst2]{CNRS, IRAP, 9 Av. colonel Roche, BP 44346, F-31028
Toulouse cedex 4, France}
\address[inst3]{Universit\'e de Toulouse, INPT-ENSEEIHT-IRIT, France}
\address[inst4]{CNRS-IRIT, France}
\address[inst5]{Lawrence Berkeley National Laboratory, Berkeley
CA94720, USA}
\cortext[cor1]{lbouchet@irap.omp.eu}

\begin{abstract}
Nowadays, analyzing and reducing the ever larger astronomical datasets
is becoming a crucial challenge, especially for long cumulated
observation times. The \INTEGRALSPI{} X/\(\gamma\)-ray spectrometer is
an instrument for which it is essential to process many exposures at
the same time in order to increase the low signal-to-noise ratio of
the weakest sources. In this context, the conventional methods for
data reduction are inefficient and sometimes not feasible at all.
Processing several years of data simultaneously requires computing
not only the solution of a large system of equations, but also the
associated uncertainties. We aim at reducing the computation time and
the memory usage.
Since the \SPI{} transfer function is sparse, we have used some
popular methods for the solution of large sparse linear systems; we
briefly review these methods. We use the Multifrontal Massively
Parallel Solver (MUMPS) to compute the solution of the system of
equations. We also need to compute the variance of the solution, which
amounts to computing selected entries of the inverse of the sparse
matrix corresponding to our linear system. This can be achieved
through one of the latest features of the MUMPS software that has been
partly motivated by this work. In this paper we provide a brief
presentation of this feature and evaluate its effectiveness on
astrophysical problems requiring the processing of large datasets
simultaneously, such as the study of the entire emission of the Galaxy.
We used these algorithms to solve the large sparse systems arising
from \SPI{} data processing and to obtain both their solutions and the
associated variances.
In conclusion, thanks to these newly developed tools, processing large
datasets arising from \SPI{} is now feasible with both a reasonable
execution time and a low memory usage.
\end{abstract}


\end{frontmatter}


\section{Introduction}

Astronomy is increasingly becoming a computationally intensive field
due to the ever larger datasets delivered by observational efforts to
map ever larger volumes and provide ever finer details of the
Universe. In consequence, conventional methods are often inadequate,
requiring the development of new data reduction techniques. The \SPI{}
X/\(\gamma\)-ray spectrometer, aboard the \INTEGRAL{} observatory,
perfectly illustrates this trend. The telescope is dedicated to the
analysis of both point-sources and diffuse emissions, with a high
energy resolution~\citep{Vedrenne03}. Its imaging capabilities rely on
a coded-mask aperture and a specific observation strategy based on a
dithering procedure~\citep{Jensen03}. After several years of
operation, it also becomes important to be able to handle
simultaneously all the data, in order, for example, to get a global
view of the Galaxy emission and to determine the contribution of the
various emission components.

The sky imaging with \SPI{} is not direct. The standard data analysis
consists in adjusting a model of the sky and instrumental background
to the data through a chi-square function minimization or a likelihood
function maximization. The related system of equations is then solved
for the intensities of both sources and background. The corresponding
sky images are very incomplete and contain only the intensities of
some selected sky sources but not the intensities in all the pixels of
the image. Hence, images obtained by processing small subsets of data
simultaneously cannot always be combined together (co-added) to
produce a single image. Instead, in order to retrieve the low
signal-to-noise ratio sources or to map the low surface brightness
``diffuse'' emissions~\citep{Bouchet11}, one has to process
simultaneously several years of data and consequently to solve a
system of equations of large dimension. Grouping all the data
containing a signal related to a given source of the sky allows to
maximize the amount of information on this specific source and to
enhance the contrast between the sky and the background.

Ideally, the system of equations that connects the data to the sky
model (where the unknown parameters are the pixels intensities) should
be solved for both source intensities and variability timescales.
This problem, along with the description and treatment of sources
variability, is the subject of another paper~\citep{PaperII}.

It is mandatory, for example when studying large-scale and weak
structures in the sky, to be able to process large amounts of data
simultaneously. The spatial (position) and temporal (variability)
description of sources leads to the determination of several tens of
thousands of parameters, if \(\sim\)6 years of \SPI\ data are
processed at the same time. Consequently, without any optimization,
the systems to be solved can exceed rapidly the capacities of most
conventional machines. In this paper we describe a technique for
handling such large datasets.


\section{Material and methods}
\label{sec:material}

\subsection{The \SPI\ spectrometer}
\label{sec:material:foundation:instrument}

\SPI{} is a spectrometer provided with an imaging system sensitive
both to point-sources and extended source/diffuse emission. The
instrument characteristics and performance can be found
in~\citet{Vedrenne03} and~\citet{Roques03}. Data are collected thanks
to 19 high purity Ge detectors illuminated by the sky through a
coded-mask . The resulting Field-of-View (FoV) is
\(\sim\)\(30^{\circ}\) and the energy ranges from 20 keV to 8 MeV. The
instrument can locate intense sources with an accuracy of a few arc
minutes~\citep{Dubath05}.


\subsection{Functioning of the ``spectro-imager'' \SPI}
\label{sec:material:foundation}

\begin{figure}[!ht]
\begin{center}
\includegraphics[width=0.48\textwidth]{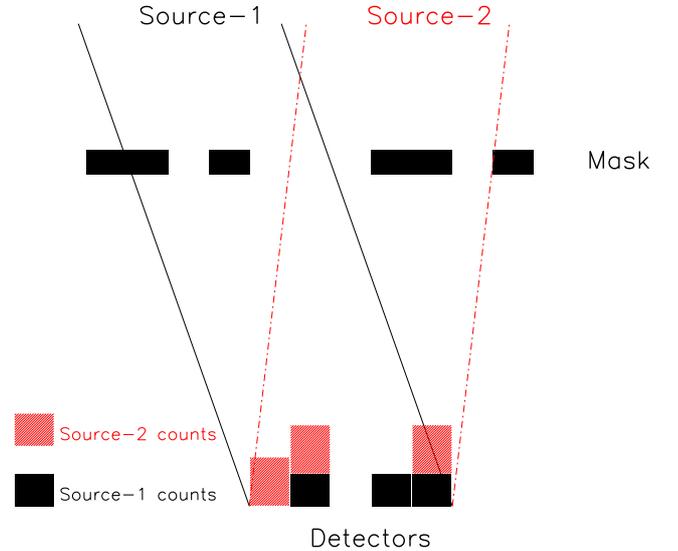}
\end{center}
\caption{SPI imaging principle. The mask consists of elements
transparent or opaque to the radiation. The opaque elements (made of
tungsten) are shown in black.The shadowgram of the mask casts onto the
detector plane (camera) depends on the source direction. Here the
counts in the different detectors of source-1 and source-2 are shown
in black and red. The counts recorded by the detectors are the sum of
all the contributions from all the sources in the FoV.}
\label{fig:shadowgram}
\end{figure}

The coded mask consists of elements which are opaque (made of
tungsten) or transparent to the radiation. Photons coming from a
certain direction cast a shadow of the mask onto the detectors plane.
The shadowgram depends on the direction of the source
(Figure~\ref{fig:shadowgram}). The recorded counts rate in each detector
of the camera is the sum of the contribution from all the sources in
the FoV. The deconvolution consists of solving a system of equation
which relates a sky model to the data through a transfer function. In
the case of \SPI, the imaging properties rely on the coded aperture,
but also on a specific observing strategy: the dithering.

\subsubsection{Dithering and sources variability}
\label{sec:material:foundation:dithering}

The reconstruction of all the pixels of the sky image enclosed in the
FoV is not possible from a single exposure. Indeed, dividing the sky
into \(\sim\)\(2^{\circ}\) pixels (the angular resolution), we obtain,
for a \(30^{\circ}\) FoV,\(\sim\)\((30^{\circ}/2^{\circ})^2=225\)
unknowns. However, a single exposure contains only 19 data values
which are the number of observed counts in the 19 Ge detectors and
does not permit us to obtain the parameters necessary to determine the
model of the sky and background. The related system of equations is
thus undetermined. The dithering observation technique aims to
overcome this difficulty.

By introducing multiple exposures for a given field that are shifted
by an offset that is small compared to the size of the FoV, it is
possible to increase the number of equations, by grouping exposures,
until the system becomes determined and thus solvable. An appropriate
dithering strategy~\citep{Jensen03} has been used where the spacecraft
continuously follows a dithering pattern throughout an observation. In
general, the pointing direction varies around a target by steps of
\(2^{\circ}\) within a five-by-five square or a seven-point hexagonal
pattern. A pointing (exposure) lasts between 30 and 60 minutes. Thus,
the dithering allows to construct a solvable system of equations.

However, in addition to the variable instrumental background, sources
are also variable on various timescales ranging from hours (roughly
the duration of an exposure) to years. This is not a major problem at
high energy (E \(\gtrsim\) 100 keV), since there are only few emitting
sources, whose intensities are rather stable in time with respect to
the statistics. At lower energies (E \(\lesssim\) 100 keV) and in most
cases, the source intensities vary during the time spanned by the all
the exposures. The chi-square, of the associated least-square problem,
for this group can be relatively high, if sources intensity variations
are not taken into account. In spite of this, it is possible to
include a model of the source intensity variations in the formulation
of the problem and to re-optimize the system of equations
accordingly~\citep{PaperII}. Nevertheless, including sources
variability in the system of equations increases the number of
unknowns to determine (\ref{sec:material:foundation:problem}) since
intensities, in each ``time-bin'' (a segment of time where the
intensity of a given source does not change statistically), are to be
determined simultaneously along with the parameters which model the
instrumental background.


\subsubsection{Cases where it is better to process large amount of
data simultaneously}
\label{sec:material:foundation:largedataset}

It is impossible from a single exposure (19 data values) to obtain the
sky image in the \(30^{\circ}\) FoV; only a coarse image containing at
most 19 sources can be obtained. This coarse image is under-sampled
and contains information on only 19 pixels (there is no information on
the other pixels). Hence, images cannot always be combined together
(co-added) to produce a single image. Furthermore, crowded regions
like the Galactic Center contain hundreds of sources and thus a single
exposure cannot provide the amount of information needed, even to
build only a coarse image. The grouping of the exposures, by selecting
all those that contain signal on a specific sky target, can provide
the necessary information. The FoV spanned by these exposures is large
(\(30^{\circ}\) to \( 60^{\circ}\)) and contains numerous sources.


\subsubsection{Problem formulation}
\label{sec:material:foundation:problem}

The signal (counts and energies) recorded by the \SPI{} camera on the
19 Ge detectors is composed of contributions from each source in the
FoV plus the background. For \(n_s\) sources located in the field of
view, the data \(D^{raw}_{dp}\) obtained from detector \(d\) during an
exposure (pointing) \(p\), for a given energy band, can be expressed
by the relation:
\begin{equation}
D^{raw}_{dp}=\sum_{j=1}^{n_s} R_{dp,j} I^{s}_{p,j} + B^{bg}_{dp} +\epsilon_{dp} \label{eqn:expression}
\end{equation}
where \(R_{dp,j}\) is the response of the instrument for source \(j\)
(function of the source direction relative to the telescope pointing
axis), \(I^{s}_{p,j}\) is the flux of source \(j\) during pointing
\(p\) and \(B^{bg}_{dp}\) the background both recorded during the
pointing \(p\) for detector \(d\). \(\epsilon_{dp}\) are the
measurement errors on the data \(D^{raw}_{dp}\), they are assumed to
have zero mean, to be independent and normally distributed with a
known variance \(\sigma_{dp}\) (\(\epsilon_{dp} \sim
N(0,[\sigma_{dp}^2]) \) and \(\epsilon_{dp} = \sqrt{D^{raw}_{dp}}\)).

For a given pointing \(p\), \(D^{raw}_{dp}\), \(R_{dp,j}\), and
\(B^{bg}_{dp}\) are vectors of \(n_d\) (say \(n_d=19\)
detectors\footnote{The number of functioning detectors could be
\(n_d=16, 17, 18\) or 19 for single events and up to 141, when all the
multiple events are used in addition to the single
events~\citep{Roques03}.}) elements. For a given set of \(n_p\)
exposures, we have a system of \(n_p \times n_d\) equations
(Eq.~\ref{eqn:expression}). To reduce the number of free parameters
related to background, we take advantage of the stability of relative
counts rate between detectors and rewrite the background term as:
\begin{equation}
B_{dp}^{bg} = I^{bg}_p \times U^d \times t_{dp} \label{eqn:backmodel}
\end{equation}

where \(I^{bg}_p\) is a normalization coefficient per pointing related
to the background intensity, \(U^d\) is a background count rate
pattern (uniformity map) on the \SPI{} camera for detector \(d\), and
\(t_{dp}\) the effective observation time for pointing \(p\) and
detector \(d\). The number of parameters necessary to model the
background reduces to \(n_p\) if \(U\) is assumed to be
known\footnote{Derived from ``empty-field''
observations~\citep{Bouchet10}.}. However, in some cases it can be
determined while processing the data (\ref{app:two:autoflfi}). \\
The two extreme cases, in terms of number of parameters to be determined, are

\begin{itemize}
\item First, when the sources and background intensities are assumed
to be constant throughout all the observation (time spanned by the
exposures), the relation between the data and the sky model can be
written, omitting the detector indices, as 
\begin{equation}
\begin{split}
               & D^{raw}_p=\sum_{j=1}^{n_s} R_{pj} I^{s}_j +  P_p I^{bg}  +\epsilon_p \\
\mathrm{with~} & P_p=t_{dp}\times U^d
\end{split}
\end{equation}
The aim is to compute the intensities \(I^{s}(j)\) of the \(n_s\)
sources and the background relative intensity \(I^{bg}\). Therefore,
the relation can be written in matrix form, as
\[
\left( {\begin{array}{cccc}
 P_1    & R_{1,1} & ...    & R_{1,n_s} \\
 P_2    & \ddots & \ddots & R_{2,n_s} \\
\vdots  & \ddots & \ddots & \vdots   \\
P_{n_p} & ...    & ...    & R_{n_p,n_s} 
 \end{array} } \right)
\left( {\begin{array}{c}
 I^{bg}   \\
 I^{s}_1 \\
 I^{s}_2 \\
 \vdots   \\
 I^{s}_{n_s} 
 \end{array} } \right)
=
\left( {\begin{array}{c}
 D^{raw}_1     \\
 D^{raw}_2     \\
\vdots          \\
D^{raw}_{n_p -1} \\
D^{raw}_{n_p}
 \end{array} } \right)
+
\left( {\begin{array}{c}
 \epsilon_1     \\
 \epsilon_2     \\
\vdots          \\
\epsilon_{n_p -1} \\
\epsilon_{n_p}
 \end{array} } \right)
\]
We can rewrite the system in a more compact form as
\begin{equation}
  y=H_0 x +\epsilon \mathrm{~or~} y_i=\sum_{j=1}^{n_s+1} h_{ij} x_j + \epsilon_i \mathrm{~for~} i=1,..,M
  \label{eqn:h0}
\end{equation}
where \(H_0\) (elements \(h_{ij}\)) is an \(M \times (n_s+1)\) matrix
and \(M=n_d \times n_p\). The parameters to be determined,
\(x=(I^{bg},I^{s}_1, \cdots, I^{s}_{n_s})\) is a  vectors of length
\(n_s+1\). The data \(y=(D^{raw}_1,D^{raw}_2,\cdots, D^{raw}_{n_p})\)
and the associated statistical errors
\(\epsilon=(\epsilon_1,\epsilon_2, \cdots,\epsilon_{n_p})\) are
vectors of length \(M\).
\item Second, if the background or the sources are variable on the
exposure timescale, the number of unknowns (free parameters) of the
set of \(n_p \times n_d\) equations is then \((n_s+1) \times n_p\)
(for the \(n_s\) sources and the background intensities, namely
\(I^{s}\) and \(I^{bg}\)).This leads, unless the number of sources is
small, to an underdetermined system of equations. \footnote{With the
Compressed Sensing approach (\cite{Bobin08, Wiaux09} and references
therein), it is possible to find a sparse solution even if the system
is underdetermined and for systems in which the matrix is sparse.}
\end{itemize}
Fortunately, in real applications, many sources vary on timescales
larger than the exposure. This leads to a  further reduction of the
number of parameters compared to the case where all sources vary on
the exposure timescale. In addition, many point sources are weak
enough to be considered as having constant flux within the statistical
errors, especially for higher energies (E \(\gtrsim\) 100 keV). Then the \(n_p \times n_s\)
parameters related to sources will reduce into \(N_s^{eff}\)
parameters and, similarly, \(N_b\) for the background. As these
parameters have also a temporal connotation, they will hereafter be
referred to as \timebins{}.

If the source named or numbered \(J\) is variable, then the total
duration covered by the \(n_p\) exposures is divided into \(K_J\)
sub-intervals where the source intensity can be considered as
stable/constant regarding the data statistics. The solution \(x_J\) is
expanded in \(K_J\) segments, it takes the value \timebins{} \(s^J_k\)
in segment k, and can be written in compact notation
\begin{equation*}
x_j=\sum_{k=1}^{K_J} s^J_k I^J_k \mathrm{~with~}
\begin{cases}
I^J_k=1 &\text{ if } t \in [t^J_{k -1},t^J_{k}[\\
I^J_k=0 &\text{ otherwise}
\end{cases}
\end{equation*}
Actually the instants \(t^J_k\) correspond to the exposure acquisition
time (exposure number), with \(t_0\)=1 and \(t^J_{k}=n_p+1\). There is
 at least one and at most \(n_p\) time segments for each source \(J\)
 (\(x_J=[s^J_1,\vdots,s^J_{K_J}]\) becoming a vector of length \(K_J\)).
The matrix \(H_0\) (eq.~\ref{eqn:h0}) is to be modified accordingly.

When expanding matrix \(H_0\), column \(J\) is expanded in
\(K_J\) new columns, hence the number of intensities (unknowns)
increases. Schematically \(H_0\) (\(M \times (n_s+1)\)) is mapped into
a matrix \(H\) (\(M \times N\)), \(N\) being the sum of all sources
intervals (\(N=\sum_{J=0}^{n_s} K_J\)), that is the number of
\timebins{} (the index J=0 correspond to the background). Matrix
\(H(1:M,1:K_0)\) is related to the background while \(H(1:M,K_0+1:N)\)
is related to the sources response. Parameters \(x(1:K_0)\) and
\(x(K_0+1:N)\) are related to background and source intensity
variations with the time (number of exposures). Box I illustrates
schematically how the matrix \(H\) is derived from the matrix \(H_0\).

\begin{small}
\begin{equation*}
\begin{split}
& H_0=
\left( {\arraycolsep=0.5mm\begin{array}{ccccc}
 h_{11}  & h_{12}  & h_{13} & ...    & h_{1N}  \\
 h_{21}  & h_{22}  & h_{23} & \ddots & h_{2N}  \\
 h_{31}  & h_{32}  & h_{33} & \ddots & h_{3N}  \\
 \vdots  & \ddots  & \ddots & \ddots & \vdots \\
 h_{M1}  & h_{M2}  & h_{M3} &...    & h_{MN}  
 \end{array} } \right)\\
\longmapsto &
H= \left( {\arraycolsep=0.5mm\begin{array}{cccccccccccc}
h_{11} & 0      & 0      &   0    &  h_{12}  & 0      & 0      & h_{13} & 0      & h_{1N} & ...    & 0      \\
0      & h_{21} & 0      &   0    &  h_{22}  & 0      & 0      & 0      & h_{23} & h_{2N} & ...    & 0      \\
0      & 0      & h_{31} &   0    &  0       & h_{32} & 0      & 0      & h_{33} & 0      & ...    & h_{3N} \\
\vdots & \vdots & \vdots & \vdots & \vdots   & \vdots & \vdots & \vdots & \vdots & \vdots & \ddots & \vdots \\
0      & 0      & 0      & h_{M1} &  0       & 0      & h_{M2} & 0      & h_{M3} & 0      & ...    & h_{MN}   
 \end{array} } \right)
\end{split}
\end{equation*}
\end{small}
Finally, the relation between the data and the sky model, similarly as in eq.~\ref{eqn:h0}, is
\begin{equation}
H x=y + \epsilon
\end{equation}
Physically, \(H\) corresponds to the transfer function or matrix,
\(y\) to the data and \(x\) to the unknown intensities (sources plus
background) to be determined (a vector of length N).

Taking into account the variability of sources and instrumental
background increases the size of the system of equation and the number
of unknowns, but also increases the sparsity of the matrix related to
the system of equations, which means that the underlying matrices have
very few non-zero entries. In our application, the matrix \(H_0\) is
sparse, thus matrix \(H\) is even sparser. Objective methods to
construct the matrix \(H\) from \(H_0\) are described
in~\citet{PaperII}.

To give an idea, for the dataset which corresponds to (\(\sim\)6 years
of data, the number of free parameters \(N=N_s^{eff} + N_b\) to be
determined are between \(N\sim 5\,000\) and \(N \sim 90\,000\)
depending on the energy band considered and hypotheses made on sources
and background variability timescale (\ref{sec:material:material}).

\subsection{Material}
\label{sec:material:material}

The material is related to the analysis of data accumulated between
2003 and 2009 with the spectrometer \SPI. The astrophysical
application is the study of diffuse emission of the Galaxy. The
details can be found in~\citet{Bouchet11}. The goal is to disentangle
the ``diffuse'' emission (modeled with 3 spatially extended
components) from the point-sources emission and instrumental
background. We need to sample these large-scale structures efficiently
over the entire sky and consequently use the maximum amount of data
simultaneously, since a single exposure covers only one-hundredth of
the total sky area. The datasets consist of 38\,699 exposures that
yield \(M=649\,992\) data points. In most cases considered here, the
background intensity is considered to be quite stable on a \(\sim\)6
hours timescale, which corresponds to \(N_b \simeq~5870\) unknowns.

\begin{enumerate}[(a)]
\item The highest energy bands (\( E \gtrsim 100\) keV) are less
problematic in terms of number of parameters to determine, as
illustrated by the 200-600 keV band. The sky model contains only 29
sources which are essentially not variable in time (given the
instrument sensitivity). The number of unknowns is \(N \simeq 5900\).
\item The lowest energy bands (\( E \lesssim 100\) keV) are more
problematic. We use the 25-50 keV band. The sky model contains 257
sources variable on different timescales. When the background
intensity is assumed to vary on \(\sim\)6 hours timescale, \(N \simeq
22\,500\) \timebins\ intensity are to be determined.\\ In some
configurations, essentially used to assess the results, background
intensity and/or strongest variable sources vary on the exposure 
timescale, and the number of unknowns could be as high as \(N \simeq
55\,000\) to \(N \simeq 90\,000\). Nevertheless, the matrices
associated with these problems remain relatively structured. \item To
avoid excessively structured matrices, we generate also matrices
\(H\), with a variable number of columns, the number of segments
\(K_J\) for a particular source being a random number between 1 and
\(n_p\). This results in a different number of parameters \(N\).
\end{enumerate}
Another astrophysical application is the study of a particular source
or sky region, here the crowded central region of the Galaxy. In this
case, it is possible to use a smaller number of exposures. We use 7147
exposures which cover a sky region of radius \(30^{\circ}\) around the
Galactic center. We measure the intensity variations of a set of 132
sources. The number of parameters to determine  \(N=3\,578\) is
relatively small. Details can be found in~\citet{PaperII}. A second
matrix, used for verification purposes, has \(N =9\,437\). It
corresponds to the case where some sources are forced to vary on
shorter timescales.

The material consists of rectangular matrices \(H\) and symmetric
square matrices \(A\) (\(A=H^{T}H\)) related to the above physical
problems (\ref{sec:material:foundation:problem}). The characteristics
of some of these matrices are described in Table~\ref{table:sparsity}.

The system we use in the experiments consists of an Intel i7-3517U
processor with 8~GB main memory. We ran the experiments on a single
core, although our algorithms are amenable to parallelism.

\begin{table}[!ht]
\caption{Sparsity of matrices \(H\) and \(H^TH\).}
\resizebox{0.48\textwidth}{!}{
\centering\begin{tabular}{rrllr}
\hline\hline
\(N\)   & \(\rho(H)\)(\%) & \(\rho(A)\)(\%)  &  \\
\hline
  3578 &    2.67    & 2.96  & Central Galaxy (27-36 keV)  \\
  9437 &    1.01   & 1.05    &                             \\
  5900 &    0.12    & 0.13  & Diffuse emission 200-600 keV \\
 22503 &   0.18    & 0.28  & Diffuse emission 25-50 keV  \\
 55333 &     0.07    & 0.09  &                             \\
149526 &     0.03    & 0.04  & Simulation (25-50 keV)     \\
\hline
\end{tabular}
}
\footnotetext{}{\small
\(\rho\)(Matrix) is the so-called \emph{density} of the matrix: the ratio between the number of non-zero elements in the matrix and the total number of elements in the matrix (\(M\times N\) for \(H\) and \(N^2\) for \(A=H^TH\), where \(M\) is the number of rows of \(H\).
The matrix \(H\) arising from the diffuse emission study have \(M = 672\,495\) rows.
The number of non-zero elements is constant \(nz=27\,054\,399\) for the matrices with \(N \ge 22\,503\) corresponding to the 25-50 keV band and 
\(nz=4\,677\,821\)  for the 200-600 keV band.
The matrix with \(N = 3\,578\) has \(M=124\,921 \) rows and \(nz=11\,948\,840\) non-zero elements. 
} 
\label{table:sparsity}
\end{table}

\subsection{Methods}
\label{sec:material:methods}

The mathematical problem described in
Section~\ref{sec:material:foundation:problem} and developed in
~\ref{sec:theory:methods:lsq} requires the solution of several algebraic
equations. First, if the chi-square statistic is used, a linear
least-squares problem has to be solved to estimate the parameters of
the model. Second, elements (entries) of the inverse of a matrix have
to be computed in order to determine the error bars (variances of
these parameters). Third, in some cases, the parameters are adjusted
to the data through a multi-component algorithm based on likelihood
tests (Poisson statistics); this problem leads to a non-linear system
of equations (\ref{app:two:mle}).

These three problems can be reduced to solving a linear system with a
square matrix: a linear least-squares problem \(\min_x ||Hx-y||\) can
be transformed into a square system \(Ax=b\) by use of the normal
equations\footnote{For clarity, we omit to weight the matrix H and the
data by the inverse of the  data standard deviation, see Section~\ref{sec:theory:methods:lsq}}
(\(A=H^TH\) and \(b=H^Ty\)). Similarly, computing entries of the
inverse of a matrix amounts to solving many linear systems, as
described in detail in Section~\ref{sec:theory:mumpsinv}. For the
above mentioned non-linear problem, we chose a Newton-type method;
this involves solving several linear systems as well.  Our problems
are large, but sparse (cf. Table~\ref{table:sparsity}), which
justifies the use of sparse linear algebra techniques. In
Section~\ref{sec:theory:largesystem}, we describe how we selected a
method suitable for our application.


\subsubsection{The least-square solution (LSQ)}
\label{sec:theory:methods:lsq}

The system is, in most cases, overdetermined (there are more equations
- or measures here - than unknowns), therefore there is (generally) no
exact solution, but a ``best'' solution, motivated by statistical
reason, obtained by minimizing the following merit function, which is
the chi-square\footnote{The number of counts per detector is high
enough to use the Gaussian statistics.}:

\begin{equation}
\chi^2=\sum_{i=1}^{M} \left[ \frac{y_i-\sum_{j=1}^{N} h_{ij}x_j} {\sigma_i} \right]^2
\end{equation}

\(y=(y_1,\ldots,y_M)\) is vector of length \(M\) representing the
data, \([\Sigma]\) a diagonal matrix of order \(M\) whose diagonal is
(\(\sigma_1,\ldots,\sigma_M\)), where \(\sigma_i\) is the measurement
error (standard deviation) corresponding to the data point \(y_i\).
These quantities are assumed to be known (formally \(\sigma_i = \sqrt
{y_i}\)). \(H=h_{ij}\) is a matrix of size \(M \times N\). The
least-square solution \(x=(x_1,\ldots,x_N)\) is obtained by solving
the following normal equation:

\begin{equation}
(H^T [\Sigma^{-2}] H)x=H^T[\Sigma^{-2}] y \mathrm{~or~as~} Ax=b
\end{equation}

Once the solution has been computed, the uncertainties on the
estimated solution \(x\) are needed as well. The corresponding
variance can be obtained by computing the diagonal of \(A^{-1}\):
\begin{equation}
Var(x_i) \propto a^{-1}_{i,i} \mathrm{~where~} a^{-1}_{i,j}  \mathrm{~refers~to~} (A^{-1})_{i,j}
\end{equation}


\section{Theory}
\label{sec:theory}


\subsection{Processing large datasets: efficient solution of large sparse systems of equations}
\label{sec:theory:largesystem}

Sparse matrices appear in numerous industrial applications (mechanics,
fluid dynamics, \ldots), and the solution of sparse linear systems has
been an active field of research since the 1960s. Many challenges
still arise nowadays, because industrial applications involve larger
and larger number of unknowns (up to a few billions nowadays), and
because hardware architectures are increasingly complex (multi-core,
multi-GPU, etc.).

Exploiting sparsity can significantly reduce the number of operations
and the amount of memory needed to solve a linear system. Let us take
the example of a partial differential equation to be solved on a 2D
physical domain; the domain can be discretized on a \(k\times k\) 2D
grid and using, say, finite differences, the equation can be
transformed into a sparse linear system with \(N=k\times k\) unknowns.
Without exploiting sparsity, this system would be solved in \(O(N^3)\)
operations (using an exact method), with a memory usage in \(O(N^2)\).
It has been shown that, for this particular case, the number of
arithmetic operations can be reduced to \(O(N^{3/2})\), and space
complexity to \(O(N \log N)\) by exploiting the sparsity of the matrix
\citep{Hoffman73}.

Many methods exist for solving sparse linear systems~\citep{Duff89,
Saad96}. Two main classes can be distinguished: \emph{direct
methods}, that rely on a matrix factorization (e.g., \(A=L\,U\)), and
\emph{iterative methods}, that build a sequence of iterates that
hopefully converges to the solution. Direct methods are known to be
numerically robust but often have large memory and computational
requirements, while iterative methods can be less memory-demanding and
often faster but are less robust in general. Iterative methods often
need to be \emph{preconditioned}, i.e., to be applied to a modified
system \(M^{-1}Ax=M^{-1}b\) for which the method will converge more
easily; a trade-off has to be found between the cost of computing and
using the preconditioner \(M\) and how the preconditioner improves the
convergence. The choice of a method is often complicated and strongly
depends on the application. In our case, we choose to use a direct
method for the following reasons:
\begin{itemize}
\item Memory usage is often a bottleneck that prevents the use of
direct methods, but with the matrices arising from our application,
direct and iterative methods have roughly the same memory footprint.
This is explained in the next section.
\item The matrices from our application are numerically challenging;
we found that unpreconditioned iterative methods (we tried GMRES) have
difficulties converging and that a direct method that does not
implement robust numerical features is also likely to fail (we
illustrate this in Section~\ref{sec:results}).
\item We need to compute error bars, which amounts to solving a large
number (\(O(N)\)) of linear systems with different right-hand sides
but the same matrix. This is particularly suitable for direct methods;
indeed, once the matrix of the system is factored (e.g., \(A=L\,U\)),
the factors can be reused to solve for different right-hand sides at a
relatively inexpensive cost. We describe this in
Section~\ref{sec:theory:mumpsinv}.
\end{itemize}

In this work, we use the \emph{MUMPS} (Multifrontal Massively Parallel
Solver) package. MUMPS~\citep{Amestoy01, Amestoy06} aims at solving
large problems on parallel architectures. It is known to be very
robust numerically, by offering a large variety of numerical
processing operations, and provides a large panel of features. In the
following section, we briefly describe how sparse direct methods
work. We introduce the basic material needed to understand the
algorithm used for the computation of error bars (described in
Section~\ref{sec:theory:mumpsinv}).


\subsection{Sparse direct methods}
\label{sec:theory:largesystem:sparsedirect}

Direct methods are commonly based on Gaussian elimination, with the
aim to factorize the sparse matrix, say \(A\), of the linear system
\(Ax=b\) into a product of~``simpler''~matrices called \emph{factors}.
Typically, \(A\) can be factored into \(A=L\,U\) where \(L\) and \(U\)
are lower and upper triangular matrices respectively, or
\(A=L\,D\,L^T\), where \(D\) is a diagonal matrix if \(A\) is
symmetric (which is the case in our application).

Sparse direct methods depend on the non-zero pattern of the matrix and
are optimized in that sense; specialized mathematical libraries for
tridiagonal, banded, cyclic matrices are common. If the pattern is
more complex, then the method usually consists of three phases:
\emph{analysis}, \emph{factorization} and \emph{solution}.


\subsubsection{Analysis}
\label{sec:theory:largesystem:sparsedirect:analysis}

The \emph{analysis} phase applies numerical and structural
preprocessing to the matrix, in order to optimize the subsequent
phases. One of the main preprocessing operations, called
\emph{reordering}, aims at reducing the \emph{fill-in}, i.e., the
number of non-zero elements which appear in the factors but do not
exist in the initial matrix; this step consists in permuting the rows
and columns of the initial matrix in such a way that less fill-in will
occur in the permuted matrix. Table~\ref{tab:fillin} shows the amount
of fill-in for different problems coming from our astrophysical
application when the matrices are permuted using the nested-dissection
method. For each matrix, the number of non-zero elements in the
original matrix \(A\) and in the \(L\) factor of the \(L\,D\,L^T\)
factorization of \(A\) are reported. Note that in our application, the
fill-in is not very large: the number of non-zero elements in the
factors is of the same order of magnitude as in the original
matrix. As a result, the use of sparse, direct methods is likely to
provide a good scalability with respect to the size of the matrix
produced by the application. Moreover, this implies that, for our
application, direct and iterative methods will have roughly the same
memory requirements; indeed, in an unpreconditioned iterative method,
the memory footprint is mainly due to the storage of the matrix \(A\),
while the major part of memory requirements of direct methods comes
from the factors. Note that, while our application exhibit low amount
of fill-in, this not the case in other applications; in many problems,
especially those involving PDEs on 3D physical domains, the number of
non-zero coefficients in the factors can be as big as one hundred times
more than in the original matrix. In this case, using an iterative
method can be interesting memory-wise.

\begin{table}[!ht]
\resizebox{0.48\textwidth}{!}{
\centering\begin{tabular}{lrrrrr}
\hline\hline
Matrix size &   3578 &    9437 &   22503 &   55333 &   149526 \\ 
\hline
\(nz(A)\)   & 378475 &  932143 & 1436937 & 2705492 &  9379127 \\ 
\(nz(L)\)   & 519542 & 1380444 & 2885821 & 9189447 & 14432264 \\ 
\hline
\end{tabular}
}
\caption{Number of non-zeros in the original matrix \(A\) and in the \(L\)
factor of the \(A=L\,D\,L^T\) factorization for different problems of our
experimental set.}
\label{tab:fillin}
\end{table}

\begin{figure}[!ht]
\begin{center}
\subfigure[Factors \(L+L^T\).]{\includegraphics[width=.20\textwidth]{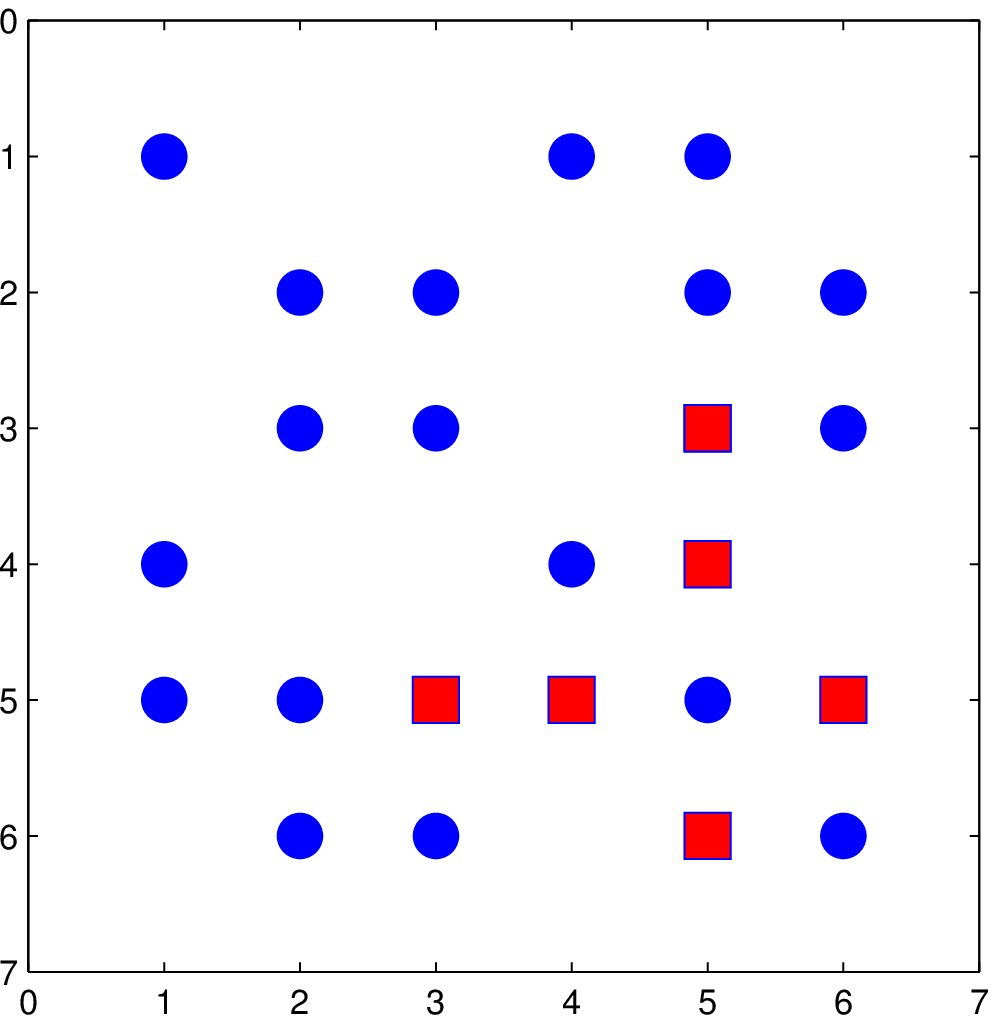} \label{fig:patternLU}}
\hfill
\subfigure[Elimination tree of \(A\).]{\includegraphics[width=.20\textwidth]{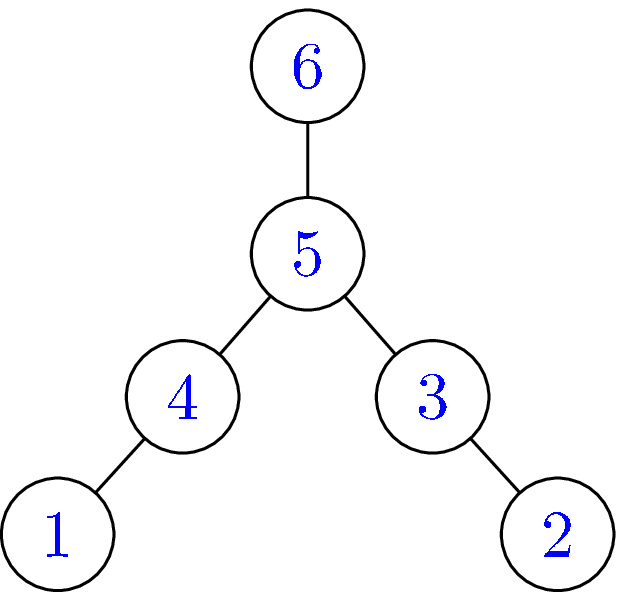} \label{fig:etreeA}}
\end{center}
\caption{The factors and the elimination tree of a symmetric matrix
\(A\). (a) pattern of the \(L+L^T\) factors of \(A\) with filled-in
entries shown with squares, (b) the elimination tree of \(A\) where
the children of a node are drawn below the node itself.}
\end{figure}

An important step of the analysis phase is the \emph{symbolic
factorization}: this operation computes the
non-zero pattern of the factors, on which the numerical factorization
and the solution will rely. The symbolic factorization computes the
structure of the factors by manipulating graphs, and also a structure
called the \emph{elimination tree}, a tree-shaped graph with \(N\)
vertices. This tree represents tasks dependencies for the
factorization and the solution phases.  We describe in more details
the elimination tree since it is a key structure to explain and
understand (see Section~\ref{sec:theory:mumpsinv}) how to accelerate
the solution phase since computing entries in the inverse of the
matrix corresponds to incomplete traversals of the elimination tree.
Figure~\ref{fig:etreeA} shows an elimination tree and we use it to
illustrate some definitions: one of the nodes is designated to be the
\emph{root}; in the figure, this is node \(6\). For our purpose, the
root is the node corresponding to the variable of the linear system
that is eliminated last. An \emph{ancestor} of a vertex \(v\) is a
vertex on the path from \(v\) to the root. The \emph{parent} (or
\emph{father}) of \(v\) is its first ancestor; all the nodes but the
root have a parent. For example, on the figure, nodes \(6\) and \(5\)
are ancestors of \(4\); \(5\) is the parent of \(4\). A \emph{child}
of a vertex \(v\) is a vertex of which \(v\) is the parent. For
example, \(4\) and \(3\) are the children of \(5\). A vertex without
children is called a \emph{leaf}; \(1\) and \(2\) are
leaves. \emph{Descendants} of a vertex \(v\) are all the nodes in the
subtree rooted at \(v\); for example, \(1\), \(2\), \(3\) and \(4\)
are descendants of \(5\).

In the following subsections (\emph{factorization} and \emph{solution
  phase}), we describe briefly how a sparse direct solver uses
elimination trees; we will also rely on this notion in
Section~\ref{sec:theory:mumpsinv} for the computation of error
bars. Further details about the construction and the role of
elimination trees in sparse solvers are given in~\citet{Liu90}.

\subsubsection{Factorization}
\label{sec:theory:largesystem:sparsedirect:factorization}

After the preprocessing performed during the analysis phase, the
numerical factorization takes place and the matrix \(A\) is
transformed into a product of factors (e.g., \(L\,U\)). The
factorization consists in traversing the elimination tree following a
\emph{postorder}, that is a \emph{topological ordering} (i.e. each
parent is visited after its children) where the nodes in each subtree
are visited consecutively. In Figure~\ref{fig:etreeA}, 1-4-2-3-5-6 is,
for example, a postorder. At each node, a partial factorization of a
dense matrix is performed. Note that nodes that belong to different
branches can be processed independently, which is especially useful in
a parallel setting.

The factorization phase tries to follow as much as possible the
preparation from the analysis phase, but sometimes, because of
numerical issues (typically, division by a ``bad pivot'', i.e. a very
small diagonal entry that could imply round-off errors), it has to
adapt dynamically: the structure of the factors and the scheduling of
the tasks can be modified on the fly.


\subsubsection{Solution phase}
\label{sec:theory:largesystem:sparsedirect:solution}

Once the matrix has been factored, the linear system is solved. For
example, in the case of the \(L\,U\) factorization, the system
\(Ax=b\) becomes \(L Ux=b\) and is solved in two steps (two solutions
of triangular systems):

\[\begin{cases}
z = L^{-1} b & \mbox{ ``Forward substitution''}\\
x = U^{-1} z & \mbox{ ``Backward substitution''}
\end{cases}\]

The forward substitution follows a bottom-up traversal of the
elimination tree as in the factorization, while the backward
substitution follows a top-down traversal of the tree. At each node,
one component of the solution is computed, and some updates are
performed on the dependent variables (ancestor nodes for the forward
phase, descendant nodes for the backward phase).


\subsection{Computing error bars: partial computation of the inverse
of a sparse matrix}
\label{sec:theory:partialinv}

In our astrophysical application, once the solution, either for the
linear or non-linear problem, has been found, it is necessary to
compute the variances of the parameters of the fitted function. In the
case of multiple regressions such as least squares problems, the
standard deviation of the solution can be obtained by inverting the
Hessian or covariance matrix. However, since the inverse of a sparse
matrix is structurally full, it is impractical to compute or store
it~\citep{Duff88}. In our case, the whole inverse of the covariance
matrix is not required: since we only want the variances of the
parameters (not their covariances), we only need to compute the
diagonal elements of the inverse.

Some work has been done since the 1970s in order to compute a subset
of elements of the inverse of a sparse matrix. One of the first works
is \citet{Takahashi73} which has been extended in~\citet{Campbell95};
this approach relies on a direct method (i.e. on a factorization). An
iterative method has been proposed in~\citet{Tang09} for matrices with
a decay property. Some methods have also been developed for matrices
arising from specific applications; a more detailed survey is given in
\citet{Amestoy10}. Many of these methods provide sophisticated ideas
and interesting performance on specific problems, but no software
package is publicly available, with the exception of the approach
implemented within MUMPS solver, that we describe in the next section.


\subsubsection{MUMPS \(A^{-1}\) feature}
\label{sec:theory:mumpsinv}

The \(A^{-1}\) feature in MUMPS has been described in
\citep{Slavova09} and was motivated by the \INTEGRALSPI{} application,
among other applications that require the computation of inverse
entries, or, more generally, applications that involve sparse
right-hand sides (as explained in this section). This feature is able
to compute any set of entries of \(A^{-1}\), relying on a traditional
solution phase, i.e. by computing every required entry \(a_{ij}^{-1}\)
as:

\[a_{ij}^{-1} = \left(A^{-1}e^{\dag}_j\right)_i\]

Using the \(LU\) factors of \(A\), this amounts to solving two
triangular systems:
\[\begin{cases}
Lz & =e^{\dag}_j\\
a_{ij}^{-1} &= \left(U^{-1}z\right)_i
\end{cases}\]

The first triangular system in the equation above is particular
because its right-hand side \(e^{\dag}_j\) is very sparse (only one
non-zero entry). Furthermore, we do not need the whole solution of the
second triangular system, but only one component. This information can
be exploited to reduce the traversal of the elimination tree; while a
regular solution phase would consist in visiting the whole elimination
tree twice (a bottom-up traversal followed by a top-down traversal),
computing \(a_{ij}^{-1}\) consists in two partial traversals of the
tree: the first triangular system is solved by following the path from
node \(j\) to the root node, and the second triangular system is
solved by following the path from the root node to node \(i\); this is
referred to as \emph{pruning} the elimination tree. Since each node of
the tree corresponds to operations to be performed (arithmetic
operations, or expensive accesses to the factors in the out-of-core
case), this leads to significant improvements in computation time.
Moreover, since we do not have to manipulate dense solution vectors,
this also leads to significant savings in memory usage.

\begin{figure}[!ht]
\centering
\includegraphics[width=0.25\textwidth]{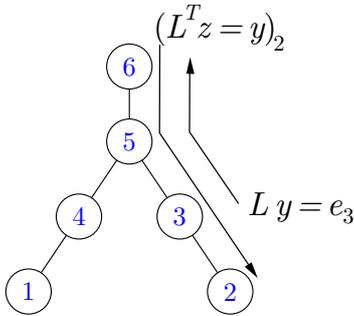}
\caption{Computation of \(a_{23}^{-1}\). The traversal of the tree is reduced
to the path from \(3\) to \(6\) and the path from \(6\) to \(2\); no
computation is performed at nodes \(1\) and \(4\).}
\label{fig:ainv23}
\end{figure}

We illustrate this technique in Figure~\ref{fig:ainv23}: entry
\(a_{23}^{-1}\) is required, thus the only nodes of tree that have to
be visited lie on the path from node \(3\) to the root node (\(6\))
and on the path from the root node to node \(2\). Therefore, one does
not have to perform operations at nodes \(4\) and \(1\).

When many entries of the inverse are requested, they cannot generally
be computed all at once (mainly because of memory usage), but they can
be computed by blocks, which allows to take advantage of efficient
dense linear algebra kernels. Work has been performed in order to find
optimal ways to form the blocks in different contexts
\citep{Amestoy10} and to improve the parallel efficiency.


\section{Calculation}
\label{sec:calculation}

A substantial time is spent in computing \(A=H^TH\) with a basic
algorithm. The use of an appropriate algorithm to perform the
operation \(A=H^TH\) helps to reduce the computation time (see
Section~\ref{sec:calculation:hess}). The MUMPS solver is used to
solve the system of equations as described in
Section~\ref{sec:calculation:mumpspackage}. Finally, the error bars
on the solution are computed, which means the calculation of the
diagonal elements of inverse matrix. The new \(A^{-1}\) feature of
MUMPS is compared with several algorithms, in terms of computation
time in Section~\ref{sec:calculation:errbars}.


\subsection{Improvements of the computation of \(A=H^TH\)}
\label{sec:calculation:hess}

The computation of the normal equation $A=H^TH$ is of paramount
importance in many problems, yet is a very challenging operation due
to the considerable amount of symbolic operations needed to compute
the sparsity structure of $A$. For this reason efficient algorithms
have been developed in the past. To perform this operation we decided
to use part of a larger code developed by~\citet{Puglisi93} for
computing the QR factorization of sparse matrices. The used part was
originally developed to compute only the structure of the $A$ matrix
and, thus, we had to extend it in order to compute the coefficients
values. This was possible thanks to the help of the original code
developer.

One important feature of this code offers the possibility to update
the elements of \(A\) that are changed after modification of some
numerical values of the columns of \(H\) without recomputing the whole
matrix (the technique used to compute simultaneously the solution and
the background pattern in the algorithm is described in
~\ref{sec:material:foundation:problem} and ~\ref{app:two:autoflfi}).

\begin{table}[!ht]
\caption{\label{table:simpleops}Time for the computation of \(A=H^TH\)}
\begin{minipage}{\linewidth}
\centering
\begin{tabular}{lrrr}
\hline\hline
Matrix                  & \multicolumn{1}{c}{Improved} & \multicolumn{1}{c}{Simple} \\
\(A=H^TH\)                & algorithm\(^a\) & algorithm\(^b\) \\
\hline
\multicolumn{3}{c}{\(N=22\,503\)} \\
Full matrix             & 28.2          & 5\,779         \\
5000 H columns modified &  0.43          &   258         \\
\multicolumn{3}{c}{\(N=149\,526\)} \\
Full matrix             & 41.2          & 18\,585         \\
5000 H columns modified &  0.13          &   31.7         \\
\hline
\end{tabular}
\end{minipage}
\footnotetext{}{ \small Times are in seconds.
  \(H\) is an \(N\) by \(M\) matrix; here \(M = 672\,495\) and \(H\) has \(27\,054\,399\) non-zero
  elements.
  \(^a\) Based on an original package from \citet{Puglisi93} and improved as suggested by the author. 
  \(^b\) \(N\) matrix vector product are used following the Compressed
  Column Storage scheme, but for each operation a dense vector of
  length \(M\) (with many zero element), that represents a column of H is
  built in place. 
}
\end{table}

Table~\ref{table:simpleops} shows the time reduction achieved for both
the computation of \(A=H^TH\) and its update after the modifications
of some columns of \(H\). The results in the first column are obtained
with the code extracted from the software package by~\citet{Puglisi93}
and improved as suggested by the author. The results on the second
column, instead, are obtained by computing \(N\) matrix vector
products where, for each product, a dense vector of length \(M\) (with
many zero elements) corresponding to a column of $H$ is built in
place.

The gain over a simple basic algorithm is significant (a factor
\(\sim\)300) and demonstrates the interest of using specialized
libraries dedicated to sparse matrix computations.


\subsection{Solving a sparse linear system}
\label{sec:calculation:mumpspackage}

Here we briefly illustrate the interest of exploiting sparsity of the
matrix when solving a linear system. In Table~\ref{table:solvetime},
we compare the time for solving linear systems arising from our
application using a dense solver (LAPACK~\citep{Andersen90}) and a
sparse solver (MUMPS). Times are in seconds and include the
\(L\,D\,L^T\) factorization of a symmetric matrix \(A\) of order \(
N\) and the computation of the solution \(x\) of the system \(Ax=b\)
(\(x\) and \(b\) are vectors of length \( N\)). In the results related
to MUMPS, the time for the analysis phase is included. In the second
row of the table, instead, the matrix is treated as dense, hence its
full storage is used and no analysis phase is performed. For the
largest two problems, the dense algorithm cannot be used as the
memory requirements are roughly 23~GB and 167~GB respectively. We can
extrapolate that on this system, the run time would be around 22 hours
for the largest problem (instead of 6.7 seconds using a sparse
algorithm). 

\begin{table}[!ht]
\caption{\label{table:solvetime}Times (in seconds) for the computation
of the solution.}
\centering\begin{tabular}{lrrrrr}
\hline\hline
Matrix size & 3578 & 9437 & 22503 & 55333 & 149526 \\
\hline
Sparse      &  0.2 &  0.7 &   1.6 &   8.0 &    6.7 \\
Dense       &  1.2 & 20.1 & 169.9 &     / &      / \\
\hline
\end{tabular}
\end{table}

The results in Table~\ref{table:solvetime} confirm that sparse, direct
solvers achieved a good scalability on the problems of our target
application whereas dense linear algebra kernels quickly exceed the
limit of modern computing platforms.


\subsection{Time to compute error bars}
\label{sec:calculation:errbars}

In this section we present experimental results related to the
computation of error bars or, equivalently, of the diagonal entries of
the inverse matrix $A^{-1}$. Our approach that relies on the pruned
tree, presented in Section~\ref{sec:theory:mumpsinv}, is compared to
the basic, left-looking approach described in \citep{Stewart98}. In
the case of a symmetric matrix, this approach computes the diagonal
entries of the inverse matrix as
\[a_{ii}^{-1}=\sum_{k=1}^{N}\frac{1}{d_{kk}}\left(l_{ki}^{-1}\right)^2\]
where we denoted with $a_{ij}^{-1}$ and $l_{ij}^{-1}$ the coefficients
of $A^{-1}$ and $L^{-1}$, respectively. This amounts to computing, one
at a time, the columns of $L^{-T}$ and then summing the corresponding
contribution onto the $a^{-1}_{ii}$ coefficients. In this algorithm,
the sparsity of the right-hand side and of the factor matrix \(L\) is
exploited but not completely, and the experimental results discussed
below show that this results in a higher execution time. Furthermore,
because of memory issues, this simple algorithm does not allow to
simultaneously compute many diagonal entries of \(A^{-1}\); clearly
this is also a limiting factor for performance. Our implementation of
this method is based on the LDL package~\citep{Davis05}. As a second
term of comparison we also provide experimental results for a brute
force approach with no exploitation of sparsity of the right-side and
solution vectors. For this purpose, we use directly the MUMPS package
and solve several systems of equations in order to compute the inverse
matrix. In addition, we analyze the influence of grouping the
computation of the diagonal entries (1 right-hand-side (RHS) at a time
or 128 at a time). 

The experimental results for the three methods described above are
reported in Table~\ref{table:computeinverse}. For the sake of this
comparison, all these methods are executed in sequential mode although
the code of the brute force approach and of the MUMPS $A^{-1}$ feature
are parallel. The experiments were carried out on the above-mentioned
system.

These results show that the brute force algorithm becomes competitive
with respect to the simple algorithm when the entries are processed by
blocks. The MUMPS~\(A^{-1}\) feature described
in~\ref{sec:theory:mumpsinv} is significantly faster than all other
approaches and the gain increases with the size of the problem.
Pruning leads to clear gains over a strict traditional solution phase.
The gain is even larger for the largest problem due to the good
scalability of the \(A^{-1}\) algorithm with respect to the problem
size. The simple, left-looking approach shows reasonable performance
for small problems, but could not be tested on our largest matrix
because numerical pivoting, not available in LDL package, is needed
during factorization to obtain an accurate solution.

\begin{table}[!ht]
\caption{Time to compute the diagonal elements of the inverse of a symmetric matrix.}
\resizebox{0.48\textwidth}{!}{
\centering\begin{tabular}{lrrrrr}
\hline\hline
Matrix size      & 3578 &  9437 &  22503 &  55333 & 149526 \\ 
\hline
Left-looking     & 28.2 & 376.1 & 2567.9 &  489.1 &      / \\ 
MUMPS (1 RHS)    & 3.77 &  38.4 &  204.1 & 1324.9 & 8230.5 \\ 
MUMPS (128 RHS)  & 1.32 &  7.34 &   45.5 &  245.6 & 2833.5 \\ 
MUMPS \(A^{-1}\) & 0.28 &   0.9 &    4.9 &   36.0 &    9.5 \\ 
\hline
\end{tabular}
}
\label{table:computeinverse}
\footnotetext{}{ \small Execution times (in seconds) for the
computation of all the diagonal entries of the $A^{-1}$ matrix with
the left-looking, brute force and MUMPS $A^{-1}$ methods. For the
brute force approach results are provided for blocks of size 1 and
128.
}
\end{table}
\begin{figure}[!ht]
\begin{center}
\includegraphics[trim=0cm 0cm 0.cm 0cm, clip=true, width=0.45\textwidth]{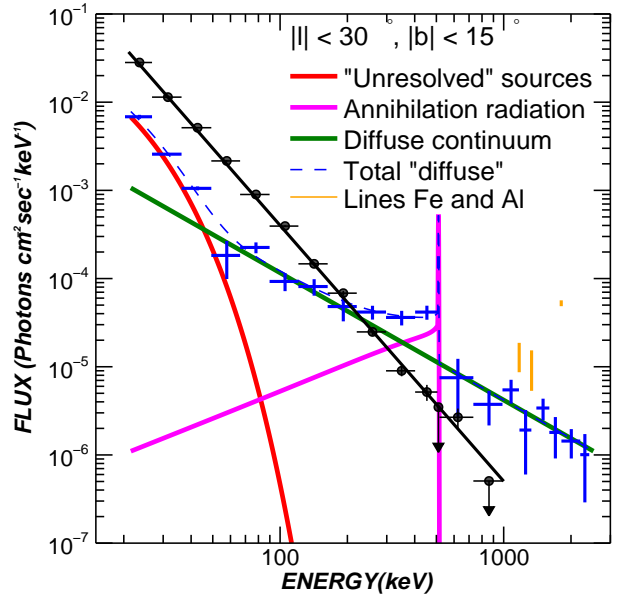}
\end{center}
\caption{Different contributions to the total emission at hard X-ray
and soft gamma-ray energies in the central radian of the galaxy. The
data points shown in black (plus filled circle) correspond to the
contribution due to 270 point sources. The average spectrum of these
sources can be viewed at http://sigma-2.cesr.fr/integral/. The data
points shown in blue correspond to the diffuse emission.}
\label{fig:diffuse}
\end{figure}
\begin{figure}[!ht]
\begin{center}
\includegraphics[trim=0.5cm 0cm 0.5cm 0cm, clip=true, width=0.48\textwidth]{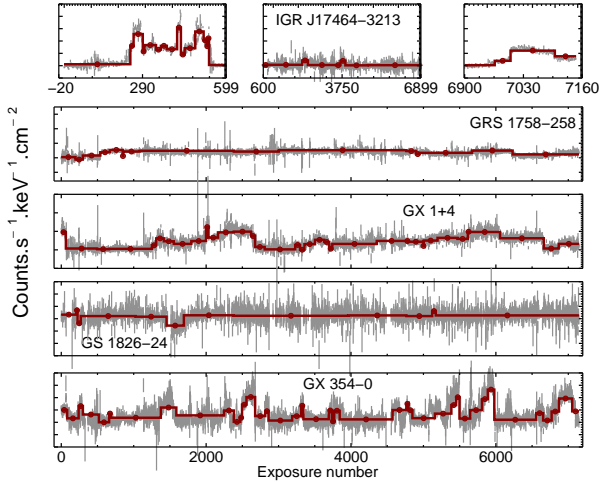}
\end{center}
\caption{Intensity evolutions (Red) of IGR~J17464-3213, GRS~1758-258,
GX~1+4, GS~1826-24 and GX~354-0 in the 27-36 keV band.  These
intensity variations are compared to the time-series (``Quick-look''
analysis) obtained with the IBIS instrument~\citep{Ubertini03} aboard
the \INTEGRAL\ observatory. The time-series (30-40 keV) is shown in
gray.}
\label{fig:datah1743fov}
\end{figure}


\section{Results and discussion}
\label{sec:results}

The MUMPS solver and its~\(A^{-1}\) functionality are the core tools
to solve systems of equations related to the measurements of the
sources intensity. Figure~\ref{fig:diffuse} shows the application to
the determination of the different components of the Galaxy spectrum.
The related analysis is performed in 24 consecutive energy bands in
order to extract counts spectra. The counts spectra are then converted
into photon spectra. The details are given in~\citet{Bouchet11}.
Another application is the study of the intensity variations of a
peculiar source or sky region. Figure~\ref{fig:datah1743fov} shows the
intensity in function of the time (exposure) of some of the sources
located in the central crowded region of the Galaxy. For this
application, the end and start of the \timebins\ are determined by a
segmentation algorithm, which is based on the efficient \(L\,D\,L^T\)
factorization of symmetric matrix provided by MUMPS, details can be
found in~\citet{PaperII}.

We have demonstrated that even for the basic operation such as sparse
matrix product, it is better to use dedicated algorithms or libraries
(\ref{sec:calculation:hess}).\\

The MUMPS solver is very effective on
the sparse matrix structure arising from astrophysical problems
encountered with \SPI.
This solver is robust and the matrix factorization is stable against
rounding errors. It provides many numerical preprocessing options and
implements robust pivoting strategies, which make it one of the most
numerically stable solvers available. The matrices arising from the
\INTEGRALSPI~ application are symmetric and indefinite; they are not

extremely challenging numerically, but they do require two-by-two
numerical pivoting for stability (in Table \ref{table:computeinverse}
the LDL package could not provide an accurate solution on our largest
matrix).  The proposed approach not only leads to substantial time
reduction but also enables the resolution of large sparse system of
equations which could not be solved using basic algorithms.

Other sparse linear systems solvers exists and have been used to
validate the performance reported in the experimental section; for an
exhaustive list see for example~\citet{Bai00}, but they all lack a
function to compute also the error bars on the solution quickly, which
is mandatory in our astrophysical application.

The \(A^{-1}\) feature in MUMPS (computation of selected inverse
entries) did not exist before the beginning of this study, the
\INTEGRALSPI{} application was actually one of the motivating
applications for developing techniques for the computation of inverse
entries, and for releasing a publicly available code. This
functionality allows to compute easily and rapidly the error bars on
the solution. The gain in time over already optimized algorithms is
important.

Among other methods to solve the problem completely, solution
and error bars, one should mention alternative methods
such as Monte Carlo Markov Chains~\citep{Metropolis53, Hastings70,
Neal93} or Simulated Annealing \citep{Kirkpatrick83}. Such advanced
statistical tools can provide the best fit and the variances of the
solution of both linear and non-linear systems of equations. In
particular MCMC methods could be useful when computing error bars, in
case of complex constraints on the function. However, these methods
may be very prohibitive in time, especially if high precision on the
parameters is required; they have in general a weak or non-guaranteed
convergence and are not the best suited for our needs, given the
complexity of our problem.


\section{Conclusions}\label{sect:Conclusions}

We have developed algorithms to process years of data and to enhance
the production of \INTEGRAL{} hard X/soft \(\gamma\)-ray survey
catalogs. These methods have been successfully applied to a set of
\(\sim\)6 years of data~\citep{Bouchet11}. 
We have shown that, for processing efficiently and accurately years of
data, it is critical to use algorithms that take advantage of the
sparse structure of the transfer function (matrix), such as those
implemented in the MUMPS software\footnote{Available at
  http://mumps.enseeiht.fr/ or http://graal.ens-lyon.fr/MUMPS/}.  It
was also demonstrated that error bars can be obtained at a relatively
inexpensive cost (the same order of magnitude as a simple problem
solution) thanks to a recently developed algorithmic feature that
efficiently computes selected entries of the inverse of a matrix. In
addition, thanks to many efforts in optimization, gains are achieved
both in memory usage and in computation time. Hence, it is possible to
process large datasets in a reasonable time and to compute the
diagonal of the covariance matrix, and thus error bars, in a rather
short time. More generally, the ideas described here can be applied to
a large variety of problems.  Finally, we are today able to solve
sparse linear systems with millions, sometimes billions, of
unknowns. Although we have not used explicitly parallel computing but
instead performed many sequential computations at the same time (for
each energy band, etc..), the proposed
approach can also be used directly in a parallel setting on massively
parallel machines.

In the near future, instruments will commonly create datasets of a few
tens to a few hundreds of Terabytes for a single observation project.
Many of the current tools and techniques for managing large datasets
will not scale easily to meet this challenge. Surveys of the sky
already require parallel computing in order to be performed. New
surveys will demand orders of magnitude increases in the available
data and therefore in data processing capabilities. It is also a
challenge for scientists who need to extract a maximum of science from
the data. Exciting scientific breakthroughs remain to be achieved as
astronomers manipulate and explore massive datasets, but they will
require advanced computing capabilities, infrastructure and
algorithms.


\section*{Acknowledgments}

The \INTEGRALSPI{} project has been completed under the responsibility
and leadership of CNES. We are thankful to ASI, CEA, CNES, DLR, ESA,
INTA, NASA and OSTC for support. We are very grateful to Chiara
Puglisi, research engineer at INPT-IRIT, for her contribution to
improve the performance of the \(H^TH\) product in
Section~\ref{sec:calculation:hess}.

\appendix


\section{Schematic solution of the system of equations}
\label{app:two}


\subsection{Maximum Likelihood Estimator (MLE) of the solution}
\label{app:two:mle}

In the case of a low number of counts, it is recommended to use the
MLE of the solution instead of the \(\chi^2\) solution.
Following~\citet{Cash79}, we maximize the likelihood function, 

\begin{equation}
L=-2 \times ( \sum_{i=1}^{M} e_i - y_i~\ln~e_i)
\end{equation}
where \(e_i\) is the model of the data obtain through the relation \(e=H x\).

\subsection{Optimization of the non-linear problem}

To optimize this non-linear problem, potentially with bound
constraints (such as positivity of the solution), there are at least
four approaches:

\begin{enumerate}[(a)]
\item Newton type methods (or modified Newton methods): they use the
Hessian matrix to define a search direction and hence need the
solution of a large linear system of equations at least at each few
iterations. They are the most powerful methods available and can find
the solution in a few iterations.
\item Quasi-Newton methods: they build an approximation of the Hessian
at each iteration. They optimize a quadratic function in at most \(n\)
iterations (\(n\) being the number of unknowns).
\item Conjugate-gradient methods: unlike the Newton-type and
quasi-Newton methods, conjugate gradients methods do not require the
storage of an \(n\) by \(n\) Hessian matrix and thus are suited to
solve large problems. The gradient of the function (a vector of length
\(n\)) is used to define the search direction. They are not usually as
reliable or efficient as Newton-type methods and often a relatively
large number of iterations has to be performed before obtaining an
acceptable solution.
\item Simplex~\citep{Nelder65}, simulated
annealing~\citep{Kirkpatrick83} or Monte Carlo Markov Chain
(MCMC)~\citep{Neal93} can also be considered, but they are often
prohibitive in time.
\end{enumerate}

Methods (a) and (b) are known as order-2 optimization methods
(gradient and Hessian used), (c) as an order-1 optimization method
(gradient used), while method (d) can use only the function value.

To use a Newton type method (order-2), we need to compute the gradient
\(G\) and the Hessian \(H_{ess}\) of the function

\begin{equation}
\begin{split}
& G_j=\frac{\partial L}{\partial x_j}= 2 \times \sum_{i=1}^{M} H_{ij} \left(1-\frac{d_i}{e_i} \right) \mathrm{~for~} j=1,..,N\\
\mathrm{and~} & H_{ess}= \frac{1}{2} \frac{\partial^2 L}{\partial^2 x}=H^T\left[\frac{d} {e^2}\right] H
\end{split}
\end{equation}

\([\frac{d} {e^2}]\) is a diagonal matrix of order \(M\) whose
diagonal is (\(\frac{d_1} {e_1^2}\),\ldots,\(\frac{d_M} {e_M^2}\)). As
for the LSQ case, the variance of the solution is obtained thanks to
the inversion of the Hessian matrix (note that in the limit \(\lim_{e
\mapsto d} ~\frac{d_i} {e_i^2} = \frac{1} {d_i}=\frac{1}
{\sigma_i^2}\), the likelihood (\(H_{ess}\)) and chi-square (\(A\)) Hessian matrices
are the same). A guess solution to this
non-linear optimization problem is the LSQ solution.

\subsection{Codes for non-linear optimization}

The fitting algorithm, based on the likelihood test statistic, is a
non-linear optimization problem. To optimize a non-linear problem,
potentially with bound constraints, a Newton type method, known for
its efficiency and reliability can be used, as we already have a
solver for large sparse systems at hand. A software package for
large-scale non-linear optimization such as IPOPT\footnote{IPOPT is
available at https://projects.coin-or.org/Ipopt} (Interior Point
OPTimizer) can be used. IPOPT uses a linear solver such as MUMPS or
MA57~\citep{Duff04} as a core algorithm. For more details on this
large-scale non-linear algorithm, see \cite{Ipotref06}. A few similar
software packages for large-scale non-linear optimization exist, among
them LANCELOT~\citep{Conn96}, MINOS~\citep{Murtagh82} and
SNOPT~\citep{Gill97}.


\subsection{``Empty-field'' auto-computation}
\label{app:two:autoflfi}

Sometimes the ``empty-field'' or ``uniformity map'' \(U\) has to be
computed with the solution. In order to preserve the linearity of the
problem, we have adopted the algorithm described below. We consider
that if the solution \(x\) is known,
\begin{equation}
y_i  =\sum_{j=1}^{K_0} h_{ij}~ x_j+ \sum_{j=K_0+1}^{N} h_{ij}~x_j =  y^B_i+y^S_i
\end{equation}
Coming back to the detector and pointing number

\begin{equation}
D^{raw}_{dp}=D^S_{dp}+U^d I_p^{bg} t_{dp} \label{eqn:A4}
\end{equation}

In the above formula \(y^{S}_{i} \equiv D^{S}_{dp}\) is the counts
due to the sources, assumed to be known.
\(y^{B}_{i} \equiv U^d I_p^{bg} t_{dp}\) is the background
contribution, \(I^{bg}\) is assumed to be known and \(U\) is to be estimated.
At this stage, using the model of the sky described by ~\ref{eqn:A4}, a rough estimate of 
the pattern is \(U^d \simeq \frac {\sum_{p=1}^{n_p}
(D^{raw}_{dp}-D^S_{dp})}{\sum_{p=1}^{n_p} t_{dp}}\).


\subsubsection{Expression for the detector pattern}

For the LSQ statistic, we wish to minimize the following quantities
for each of the working detectors,

\begin{small}
\begin{equation}
\chi^2(d)= \sum_{p=1}^{n_p} \left( \frac{D^{raw}_{dp}-D^S_{dp}-U^d I_p^{bg} t_{dp}}{\sigma_{dp}} \right)^2 \mathrm{~for~d=1,...,n_p}
\end{equation}
\end{small}

The LSQ solution \(U^{LSQ}(d)\) is 

\begin{equation}
U^{LSQ}(d)=\frac{ \sum_{p=1}^{n_p}  (D^{raw}_{dp}-D^S_{dp}) \times I_p^{bg} t_{dp} / \sigma_{dp}^2} {\sum_{p=1}^{n_p} ( I_p^{bg} t_{dp}) ^2/ \sigma_{dp}^2}
\end{equation}

For the MLE statistic, we do not have to preserve the linearity of the
problem and hence  the computation of the improved ``empty-field'' pattern can be done during the non-linear optimization
process. On another side, the algorithm is simplified if we proceed
similarly as in the LSQ case. Then, we wish to maximize the following
quantities for each of the working detectors,

\begin{small}
\begin{equation}
\begin{split}
L(d)=& -2 \left( \sum_{p=1}^{n_p} D^S_{dp}+U^d I_p^{bg} t_{dp}-D^{raw}_{dp}~\ln~[D^S_{dp}+U^d I_p^{bg} t_{dp}] \right)\\
& \mathrm{~for~d=1,...,n_p}
\end{split}
\end{equation}
\end{small}

The MLE solution \(U^{MLE}(d)\) is
\begin{equation}
U^{MLE}(d)= \frac { \sum_{p=1}^{n_p} (D^{raw}_{dp}-D^S_{dp})} {\sum_{p=1}^{n_p} I_p^{bg} t_{dp}}
\end{equation}

One should mention that it is possible to compute, similarly, an
``empty-field'' pattern on some restricted time interval instead of
the whole dataset; the best ``empty field'' for pointing intervals
\(p_k\) to \(p_{k+1}\) is then,

\begin{equation}
\begin{cases}
U^{MLE}(d,k) = \frac{ \sum_{p=p_k}^{p_{k+1}} (D^{raw}_{dp}-D^S_{dp})} { \sum_{p(k)}^{p(k+1)} I_p^{bg} t_{dp}} \\
U^{LSQ}(d,k) = \frac{ \sum_{p=p_k}^{p_{k+1}} (D^{raw}_{dp}-D^S_{dp}) \times I_p^{bg} t_{dp} / \sigma_{dp}^2} {\sum_{p(k)}^{p(k+1)} (I_p^{bg} t_{dp}^2)^2/ \sigma_{dp}^2}
\end{cases}
\end{equation}


\subsection{``Empty-field'' schematic construction}
\label{app:two:autoflfialgo}

A sub-optimal algorithm to obtain both the sources and the background fluxes, as well as the improved 
``empty-field'' pattern is described in
Algorithm~\ref{table:algo_lsq_iterback}. We start with an
approximation \(U_0\) and apply some iterative refinement. In
practice, the algorithm converges in a few iterations.

\begin{small}
\begin{algorithm}[H]
\caption{Computation of the ``Empty field'', the solution and its variance}
\label{table:algo_lsq_iterback}
\begin{algorithmic}[1]
\STATE \(U=U_0\), compute the structure of the Hessian (\(A\) or \(H_{ess}\))
\FOR[Iterative computation of U and x]{i=1 \TO itermax}
\STATE Compute LSQ or MLE solution
\STATE Compute a new approximation of \(U\) by minimizing again LSQ or maximizing MLE statistics
\STATE Update \(H\) (The first \(K_0\) columns of \(H\) and update the new Hessian matrix (Sec.~\ref{sec:calculation:hess}))
\STATE If \(\chi^2\) stops decreasing or the likelihood function stops increasing then go to step 8 
\ENDFOR 
\STATE Compute \(H\) at the solution (if not already done) and the diagonal of \(H^{-1}\) to obtain the uncertainties on the solution 
\end{algorithmic}
\end{algorithm}
\end{small}



\section{Highlights}
\begin{itemize}
\item \INTEGRALSPI{} X/\(\gamma\)-ray spectrometer data analysis
\item Large astronomical data sets arising from the simultaneous
analysis of years of data.
\item Resolution of a large sparse system of equations; solution and
its variance.
\item The Multifrontal Massively Parallel Solver (MUMPS) to solve the
equations.
\item MUMPS \(A^{-1}\) feature to compute selected inverse entries
(variance of the solution,\ldots).
\end{itemize}

\end{document}